\documentclass[useAMS,usenatbib,letterpaper]{mn2e}
\usepackage{graphicx}
\usepackage{listings}
\usepackage{amsmath}

\def\xi{\hbox{$X_{\rm i}$}}
\catcode`\@=11
\def\gsim{\ifmmode{\mathrel{\mathpalette\@versim>}}
    \else{$\mathrel{\mathpalette\@versim>}$}\fi}
\def\lsim{\ifmmode{\mathrel{\mathpalette\@versim<}}
    \else{$\mathrel{\mathpalette\@versim<}$}\fi}
\def\@versim#1#2{\lower 2.9truept \vbox{\baselineskip 0pt \lineskip 
    0.5truept \ialign{$\m@th#1\hfil##\hfil$\crcr#2\crcr\sim\crcr}}}
\catcode`\@=12
\def\msun{\hbox{$M_\odot$}}

\def\t9{\hbox{$t_9$}}

\def\mstar{\hbox{$M_\star$}}

\def\m*{\hbox{$M_*$}}

\def\ho{\hbox{$H_\circ$}}
\def\h50{\hbox{$\ho /50$}}

\def\y1{\hbox{${\rm yr}^{-1}$}}

\newcommand{\aap}{A\&A}
\newcommand{\mnras}{MNRAS}
\newcommand{\apj}{ApJ}

\newcommand{\apjs}{ApJS}
\newcommand{\araa}{ARA\&A}

\title[The dominance of quenching]{The dominance of quenching through cosmic times}
\author[Alvio Renzini]
{Alvio Renzini$^{1,2}$\thanks{E-mail: $\,$alvio.renzini@oapd.inaf.it}
\\
\\
$^{1}${INAF-Osservatorio Astronomico di Padova, Vicolo
   dell'Osservatorio 5, I-35122 Padova, Italy}\\
$^{2}$ Subaru Telescope, National Astronomical Observatory of Japan, 650 North A'ohoku Place, Hilo, Hawaii 96720, USA
}

\def\gtsima{$\; \buildrel > \over \sim \;$}
\def\ltsima{$\; \buildrel < \over \sim \;$}
\def\prosima{$\; \buildrel \propto \over \sim \;$}
\def\gsim{\lower.5ex\hbox{\gtsima}}
\def\lsim{\lower.5ex\hbox{\ltsima}}
\def\simgt{\lower.5ex\hbox{\gtsima}}
\def\simlt{\lower.5ex\hbox{\ltsima}}
\def\simpr{\lower.5ex\hbox{\prosima}}

\def\h1{$h^{-1}$}
\def\eeq{\end{equation}}
\def\beq{\begin{equation}}
\def\24mu{24\,$\mu{\rm m}$}
\def\70mu{70\,$\mu{\rm m}$}
\def\8mu{8\,$\mu{\rm m}$}
\def\msun{\hbox{$M_\odot$}}
\def\mstar{\hbox{$M_\star$}}
\def\y-1{\hbox{${\rm yr}^{-1}$}}
\def\mpc-3{\hbox{${\rm Mpc}^{-3}$}}

\begin{document}

\date{Accepted  April 5, 2016,  Received January 11, 2016 in original form}

\maketitle
                                                            
\label{firstpage}

\date{2016}

\begin{abstract}
The evolution with cosmic time of the star formation rate density (SFRD) and of the {\it Main Sequence} star formation rate--stellar mass relations are two well established observational facts. In this paper the implications of these two relations combined are analytically explored, showing that quenching of star formation 
must start already at very early cosmic times and the quenched fraction then dominates ever since over the star forming one. Thus, a simple picture
of the cosmic evolution of the global SFRD  is derived, in terms of the interplay between star formation and its quenching.
\end{abstract}





\section{Introduction}
\label{sec:introduction}
A high-level description  of star formation in galaxies through cosmic times is best given by two widely used relations, namely  the main sequence (MS) of star-forming galaxies, i.e., the tight relation between star formation rate (SFR) and stellar mass ($\mstar$, \citealt{Daddi07,Noeske07,Elbaz07}) and the so-called Lilly-Madau plot giving the time (redshift) evolution of the SFR density (SFRD) of the Universe \citep{lilly96,madau96}. Yet, how the two relations interact with each other has not been duly explored so far, an issue  that this paper is meant to address in the simplest possible fashion. 

Over the twenty  years since its discovery, much progress has been achieved in observationally establishing the evolution of the SFRD, a progress that is illustrated in the recent review by \cite{md14}, from which we adopt:
\begin{equation}
{\rm SFRD}(z) = 0.015 {(1+z)^{2.7}\over 1+[(1+z)/2.9]^{5.6}}\quad \msun\y-1\mpc-3,
\label{eq:sfrd}
\end{equation}
holding all the way from $z\simeq 0$ to $z\simeq 8$.

At the same time, much effort has been devoted to explore the evolution of the MS relation in slope, shape and normalization (e.g., \citealt {Pannella09, Peng10, 
Rodighiero11, Rodighiero14, Karim11, Popesso11, Popesso12, Wuyts11, Whitaker12, Whitaker14, Sargent12, 
Kashino13, Bernhard14, Magnelli14,renzini15,salmon15}), with the MS relation being in place  at least up to $z\sim 6$ \citep{Speagle14}. However, the MS slope and shape may differ
appreciably from one of such studies to another, depending on how star forming galaxies (SFG) are selected and how SFRs and stellar masses  are measured
(see e.g., the compilation by \citealt{Speagle14}). Following \cite{Peng10} we adopt for the specific SFR (sSFR=SFR/$\mstar$):
\begin{equation}
{\rm sSFR}(t) = 2.5\times 10^{-9}\eta\,(t/3.5\;{\rm Gyr})^{-2'2}\quad {\rm Gyr}^{-1}
\label{eq:ssfr}
\end{equation}
where  the parameter $\eta$ has been  introduced to compensate --if necessary-- for a possible mismatch resulting from different systematics in mass and SFR measurements affecting  the two relations above. Equation \eqref{eq:ssfr} holds for cosmic time $t\ge 3.5$ Gyr (i.e., $z\lsim 2$), whereas the run of sSFR for earlier times is more uncertain. For example, \cite{gonzalez10} found a flat sSFR for $z>2$, while according to \cite{stark13} the sSFR keeps increasing by another factor of $\sim 5$ up to at least $z\sim 7$, see Figure 13 in \cite{md14}.
In the following, for $t\le 3.5$ Gyr  we adopt an analytical extension of Equation \eqref{eq:ssfr}, continuous with its derivative, that gradually flattens, doubling to $5\times 10^{-9}$ G$\y-1$ by
$t=0.627$ Gyr, i.e.,  the age of the Universe at $z=8$ in the adopted cosmology, $\Omega=0.3,\; \Lambda=0.7$ and $\ho=70$.  In any way, we shall briefly comment on what would have resulted if adopting either the \cite{gonzalez10} or the \cite{stark13} recipes. In the sequel, referring to Equation \eqref{eq:ssfr} i
will mean to include this early-epoch extension.

Adopting a sSFR$(t)$ independent of mass, i.e. SFR $\propto\mstar$ as in Equation \eqref{eq:ssfr}, needs to be justified. It is indeed well known that in most determinations the MS slope tends to be lower than unity (typically $\sim 0.8$) and may flatten for high $\mstar$ values, see references above. However, \cite{Abramson14} find the slope to be  about unity if considering only the stellar mass of the star-forming portion of galaxies, having decomposed local galaxies in their active star-forming  disk and quenched bulge. Similarly, \cite{Salmi12} find a MS with slope $\sim 1$ at $z\simeq 1$ when restricting to disk-dominated galaxies (i.e., those with S\'ersic index $n<1.5$). At higher redshifts ($<\! z\! >\simeq 2.2$), the apparent bending of the MS disappears when removing from the mass budget  the mass of  (almost) quenched bulges \citep{tacchella15}. Thus, there is observational evidence supporting the notion of a linear relation between SFR and the stellar mass  of  the {\it actively star-forming portions of galaxies,} such as their {\it disks.} Therefore, here we assume that Equation \eqref{eq:ssfr} represents such relation. 

Simplicity is an additional  reason to adopt a linear MS relation between SFR and $\mstar$. The procedures to be presented in this paper could be generalized to accommodate a MS relation of any kind in slope and shape, but this would come at the expenses of simplicity, requiring a much more laborious procedure including   the mass function of galaxies and its evolution. In the present approach, we refrain from embarking in such complexities, opting instead for simplicity, with the aim of capturing just the essentials of the global evolution. 

\section{Stellar mass growth and quenching}

A first adjustment between Equation \eqref{eq:sfrd} and Equation \eqref{eq:ssfr} is required, as \cite{md14} used a Salpeter IMF while \cite{Peng10} used a Chabrier IMF. Thus, in the sequel we use a SFRD$(t)$ reduced by a factor 1.7 compared to Equation \eqref{eq:sfrd}.
With the resulting  SFRD$(t)$,  the growth of the stellar mass density of the Universe is then given by:
\begin{equation}
\rho_\star(t)=(1-R)\int_{t_\circ}^{\, t}\!\!{\rm SFRD}(t)dt,
\label{eq:rhostar}
\end{equation}
where $R$ is the return fraction due to stellar mass loss (we assume $R=0.3$) and $t_\circ=0.627$ Gyr is the cosmic time  corresponding to $z=8$. 

\begin{figure}
\vskip -2.1 cm
 \centering
 \includegraphics[width=0.52\textwidth, keepaspectratio]{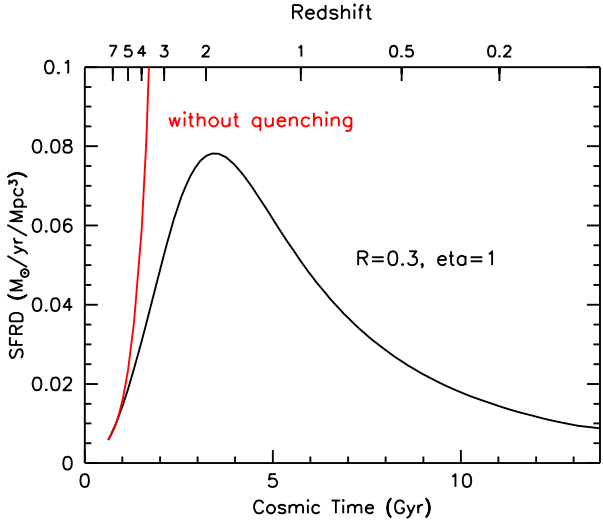}
 \vskip -2.7 cm
\caption{The run of the star formation rate density (SFRD, black line) as a function of cosmic time as from Madau \& Dickinson (2014). The red line represents the growth of the SFRD that would result from the adopted sSFR from Equation (7) in absence of quenching.}
\label{sf}
 \end{figure}

Given the stellar density and the sSFR, the SFRD of the Universe follows naturally as:
\begin{equation}
{\rm SFRD}(t) = {\rm sSFR}(t)\, \rho_\star^{\rm \! SF}(t),
\label{eq:sfd}
\end{equation}
where $\rho_\star^{\rm \! SF}(t)$ is the stellar density of those galaxies (or part of) that are actively star forming and on average follow the MS relation as given by Equation \eqref{eq:ssfr}.  Therefore, the  stellar mass in galaxies (or part of)  whose  sSFR is $\ll$ than given by Equation \eqref{eq:ssfr} is not  included in 
$\rho_\star^{\rm \! SF}$. Such non-star-forming stellar mass is  hosted by quenched galaxies, or galaxies well underway to be quenched, and by the quenched
spheroid (bulge and halo) of galaxies still hosting a star-forming disk. Thus,
\begin{equation}
\rho_\star^{\rm \! SF}(t) = {{\rm SFRD}(t)\over{\rm sSFR}(t)},
\label{eq:sfd}
\end{equation}
can be regarded as the definition of the actively star-forming, stellar mass density of the Universe. 

By the same token,
\begin{equation}
\rho_\star^{\rm \! Q}(t) = \rho_\star(t) - \rho_\star^{\rm \! SF}(t)
\label{eq:qd}
\end{equation}
is the stellar mass density of the Universe in quenched galaxies or quenched fraction of star-forming ones (e.g., quenched bulges and halos).
In other words, a stellar mass density $\rho_\star^{\rm \! SF}$  is sufficient to produce the observed SFRD, if forming stars at the main sequence sSFR as given by Equation \eqref{eq:ssfr}. Correspondingly, a stellar mass $\rho_\star^{\rm \! Q}$ must be {\it quenched}, i.e., with a much lower sSFR. 
For the sake of clarity, in this paper by quenching one means the drop of the sSFR of galaxies from its MS value  given by Equation  \eqref{eq:ssfr} to much lower values, and quenched galaxies, or portion of, are those which sSFR is $\ll$ than given by Equation  \eqref{eq:ssfr}, without any implication on the physical processes that may have led to such low sSFR.

If all the stellar mass produced by star formation were to remain actively star forming and lying on the MS with a sSFR from Equation \eqref{eq:ssfr}, i.e., in absence of any quenching, then the stellar mass would grow as:
\begin{equation}
{d\rho_\star(t)\over dt}= (1-R)\, {\rm sSFR(t)}\, \rho_\star(t),
\label{eq:grow}
\end{equation}
which --upon integration-- implies a (quasi) exponential growth of both stellar mass and SFR densities at early times (e.g., if  the sSFR is nearly constant). This is perfectly
analog to the result of  the same kind of integration once performed for individual galaxies, also giving rise to an early quasi-exponential growth of their stellar mass and SFR 
\citep{Renzini09,Peng10,leitner12,lee14,Speagle14}. The result of such integration (for $\eta=1$) is shown in Figure \ref{sf} as a red line labeled ``without quenching", whereas the black line shows the actual SFRD from Equation \eqref{eq:sfrd}, reduced by the 1.7 factor.  For a short time interval the red  line overlaps the black one, meaning that indeed at the beginning virtually all stellar mass remains star forming. But soon it  departs and keeps diverging very steeply from the actual SFRD, implying that without quenching stellar mass and SFR densities would be dramatically overproduced. This is, now on a global scale, the same  catastrophic growth that would happen to individual galaxies in absence of quenching \citep{Renzini09}. Once more, it is the high sSFR that {\it demands} quenching.  Actually, the higher the adopted sSFR the lower the $\rho_\star^{\rm \! SF}$ which is sufficient to produce the observed SFRD, and therefore the higher the resulting quenched mass $\rho_\star^{\rm \! Q}$, from Equation \eqref{eq:qd}. So, paradoxically, more quenching is required, the higher the sSFR.

\begin{figure}
 \centering
 \includegraphics[width=0.45\textwidth, keepaspectratio]{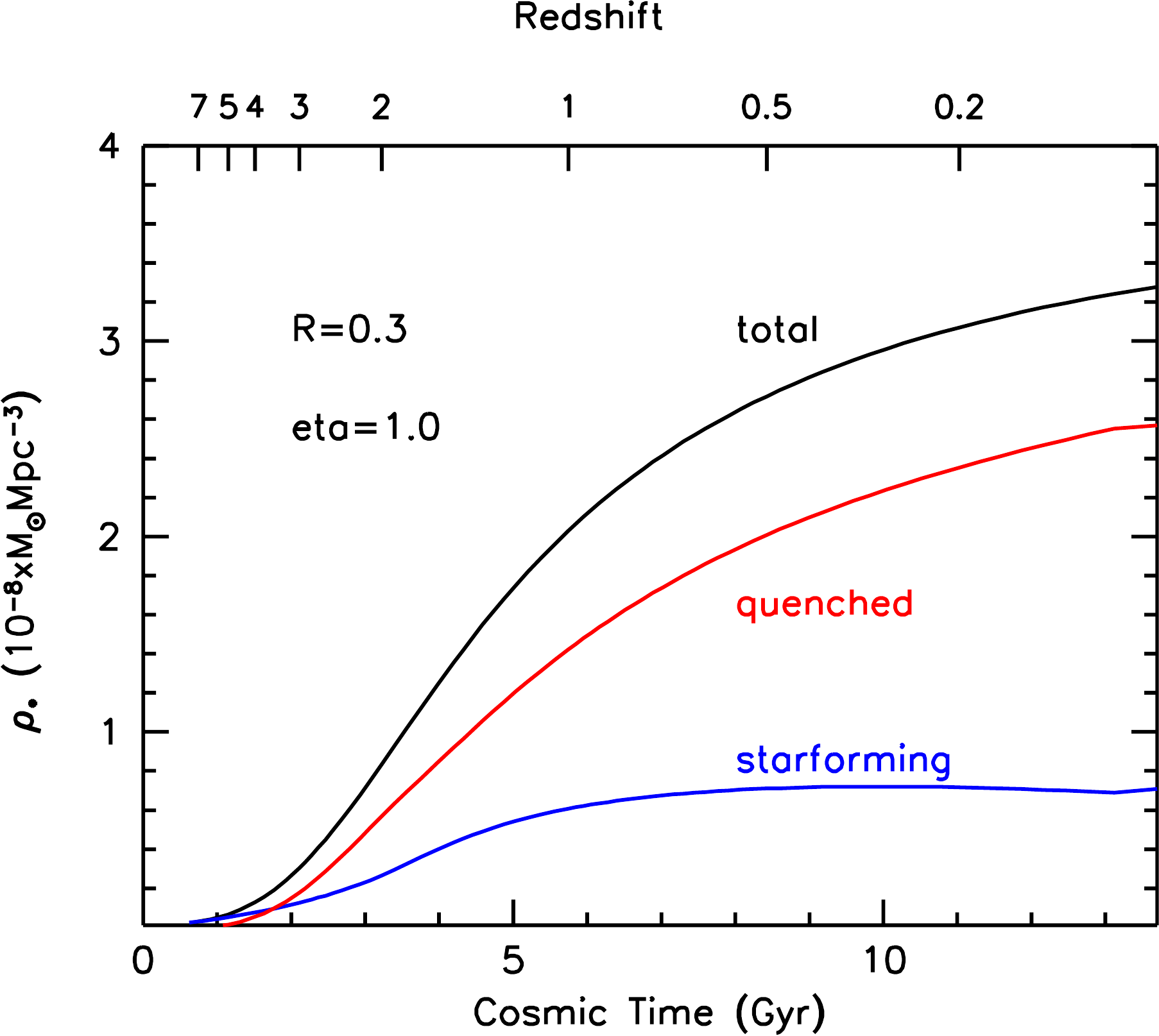}
\caption{The growth of the total stellar mass density as a function of cosmic time (black line) together with that of the quenched and star-forming fractions (red and blue lines, respectively).}
\label{rho}
 \end{figure}

Figure \ref{rho} shows the corresponding evolution of the total stellar density $\rho_\star(t)$, as well as those of the star-forming and quenched fractions,  as from Equations  \eqref{eq:rhostar},  \eqref{eq:sfd} and  \eqref{eq:qd}, respectively. Two notable results  illustrated by this figure are worth emphasizing. The first is that quenching sets in very early, and by $t=2$ Gyr (i.e., $z\simeq  3$) the quenched fraction takes over the star-forming one and increasingly dominates all the way to $z=0$. The second remarkable result is that the cosmic stellar density of the star-forming fraction remains essentially constant since $t\simeq 6$ Gyr ($z\sim 1$), and therefore the corresponding drop of the SFRD is  almost entirely due to the drop of the main sequence sSFR $\propto t^{-2.2}$ during the same time interval. 
Yet, $\rho_\star^{\rm \! SF}(t)$
remains nearly constant because the new stellar mass produced by the SFRD is almost precisely compensated by the quenching rate density, i.e., the stellar mass that quench per year and per Mpc$^3$. Thus, both quenching and the secular decline of the sSFR of MS galaxies concur in determining the cosmic decline of the SFRD since $z\sim 2$.

From Figure \ref{rho} we also see that at the present time (i.e., $t=13.7$ Gyr) the star-forming and quenched fractions account for respectively $\sim 22$\% and $\sim 78\%$ of the total stellar mass density of the Universe. This compares quite favorably with what is observed locally. From the bulge-disk decomposition operated by \cite{Abramson14} at $z\sim 0$ it is indeed estimated that only $\sim 23\%$ of the stellar mass is in star-forming environments today, i.e., in star-forming disks, the remaining fraction being in quenched galaxies and in quenched bulges (L.E. Abramson, private communication). This is also in fair agreement with the \cite{moustakas13} finding that SFGs account locally for $\sim 40\%$ of the stellar mass, which reduces to $\sim 25-28\%$ when excluding their quenched bulges from the mass budget, which accounts for $\sim 30-40\%$ of their stellar mass (e.g., \citealt{lang14}). 

\section{Caveats and discussion}
The results illustrated in Figures \ref{sf} and \ref{rho} derive uniquely from the cosmic star formation history and the evolution of the main sequence of
SFGs as given by Equations \eqref{eq:sfrd} and \eqref{eq:ssfr}, respectively. As such, the numerical results may change if in the future more accurate relations could be derived from observations. In any event, these results are more solid for $t\gsim 3.5$ ($z\lsim 2$) than at earlier times, basically because the sSFR is more uncertain  at higher redshifts, say $z>3$ ($t\lsim 2$ Gyr) when, nevertheless, only a quite minor fraction of the final stellar mass is built up. 

Assuming a sSFR$(t)$  that keeps increasing as $t^{-2.2}$ all the way to $t=t_\circ=0.627$ Gyr  would  produce a $\rho_\star$  from Equation \eqref{eq:grow} growing extremely fast at early time and a SFRD=sSFR$\times\rho_\star$ running almost vertically already since $t=t_\circ$. To avoid such overgrowth,  an  enormous quenching rate would be required to keep the SFRD at the actual observed level given by Equation \eqref{eq:sfrd}. In n practice,  the quenching rate  should almost equal  the SFRD itself, which looks quite implausible. Even the less extreme sSFR$(t)$ from \cite{stark13}, with its factor of $\sim 5$ increase beyond $z=2$, would encounter the same problem, as would the sSFR from \cite{salmon15} with it factor of $\sim 10$ increase between $z=2$ and $z\sim 7$, or the factor $\sim 8$ increase found by \cite{faisst16}.

If instead one adopts a sSFR that remains constant beyond $z\sim 2$, then the $\rho_\star^{\rm \! SF}(t_\circ)$  required to match the observed 
SFRD$(t_\circ)$ at $z\sim 8$ would be a factor $\sim 3$  higher  than  $\rho_\star(t_\circ)$ as reported by \cite{md14}. We conclude that our heuristic assumption of a smooth and flattening increase of the sSFR for $t<3.5$ Gyr does not incur in these  opposite difficulties, actually minimizes the required quenching at early times and is more keen the factor of $\sim 2$ increase from $z=2$ to  $z\sim 6$ found by \cite{gonzalez14}, see also \cite{tasca15}. In any way, the actual run of the quenched and star-forming fractions at very  early times since $t_\circ$ remains uncertain, modulo the adopted sSFR. 

However, at later times, the relative contributions of the quenched and star-forming fractions are independent of what is assumed for the sSFR at early times, hence the results are definitely  more robust. The main uncertainty remains the possibility of a relative systematics between Equation \eqref{eq:sfrd} and Equation \eqref{eq:ssfr}, as SFRs and stellar masses were derived independently by different teams. This can be quantitatively explored by varying the parameter $\eta$ introduced in Equation \eqref{eq:ssfr}. For example, if \cite{Peng10} had systematically overestimated the sSFR by, say,  20\% relative to \cite{md14}, then one should take  $\eta=0.8$ to restore consistency. Therefore, from Equation \eqref{eq:sfd} the resulting $\rho_\star^{\rm \! SF}(t)$ would be 20\% higher than shown in Figure \ref{rho} with $\rho_\star^{\rm \! Q}(t)$ being correspondingly reduced as demanded by Equation \eqref{eq:qd}. Thus, the parameter $\eta$ controls the relative proportion of $\rho_\star^{\rm \! SF}(t)$ and $\rho_\star^{\rm \! Q}(t)$, but it is indeed quite reassuring  that a  reasonable result at $z=0$ is obtained just for $\eta=1$.

A plot such as that in Figure \ref{rho} is potentially subject to observational check at all redshifts, mapping the growth of both $\rho_\star^{\rm \! SF}(t_\circ)$ and $\rho_\star^{\rm \! Q}(t_\circ)$ as they have been defined above. While this goes beyond the scope of this paper, a few considerations are in order.
\cite{muzzin13} and \cite{tomczak14} have used the rest-frame $UVJ$ selection \citep{wuyts07} to map the stellar mass density in star-forming and quenched galaxies all the way to $z\sim 4$, finding
that only at $z\lsim 1$ the mass density in quenched galaxies takes over that of star-forming ones, at apparently macroscopic variance with what shown in Figure \ref{rho}, where this happens at a much earlier time, i.e., at $z\sim 3$. However, the $UVJ$ selection  was used  to pick only fully quenched galaxies and therefore did not account for the (bulge/spheroid) quenched portion of SFGs. So, a comparison with the \cite{muzzin13} and \cite{tomczak14} results is far from being conclusive. Moreover, a bimodal distribution of galaxies in the $UVJ$ plot is recognizable only up to $z\sim 2$ and beyond this redshift the distinction between star-forming and  quenched galaxies gets more difficult.
In addition, at all redshifts it is easier to identify SFGs  rather than quenched ones and mass functions can be traced only down to an higher and higher  mass completeness limit. For these reasons, it appears quite premature to overplot  existing data on diagrams such as Figure \ref{rho}. This should  rather be taken as a predictions for future observations to check.

Critical will be a bulge-disk decomposition at all redshifts, able to estimate the sSFR in a space-resolved fashion, hence able to distinguish quenched {\it bulges} 
and their star-forming {\it disks}. Few such attempts have been made so far. For example, using H$\alpha$ emission this has been possible only for a handful of galaxies at $z\sim 2.2$ and only after much observational efforts \citep{genzel14,tacchella15}, or at $z\sim 1$ from two-dimensional HST grism spectra \citep{nelson15}.  A mere  morphological bulge-disk decomposition is indeed insufficient to distinguish the quenched and star-forming parts of high redshift galaxies (e.g., \citealt{lang14}), as star-forming bulges and quenched disks also exist. For example, the majority of quenched galaxies at $z\sim 2$ are disk dominated, with also the disk being quenched \citep{vdw11}.  According to \cite{lang14}, at $z\sim 2$  about $30\%$ of stellar mass is in fully quenched galaxies and $\sim 70\%$ is in star-forming galaxies. In these latter galaxies, bulges account for $\sim 40\%$ of the stellar mass and if one can assume that such bulges are quenched then the star forming, disk  fraction would be $\sim 40\%$ and  the corresponding quenched fraction $\sim 60\%$.
This is not too far from the   $\rho_\star^{\rm \! Q}/\rho_\star^{\rm \! SF}$ ratio $\simeq 2$ at $z=2$ ($t\simeq 3.5$ Gyr)  as shown in Figure \ref{rho}.  In 
\cite{lang14} the bulge-disk decomposition assumes that bulges and disks have S\'ersic index $n=4$ and 1, respectively. However, there cannot be a one-to-one correspondence between morphology in one specific photometric band and star formation activity (sSFR): quenched bulges can indeed have $n<4$, while other $n=4$ bulges may still be actively star  forming \citep{Wuyts11}. 
So, the above comparison between the predicted quenched fraction  and that inferred from existing observations can only be regarded as very preliminary.

What is needed is e.g., a space-resolved, rest-frame $UVJ$ mapping of galaxies able to distinguish their star-forming and quenched parts, but this  technology is not yet fully in place. Beyond the $H$ or $K$ band, existing data lack sensitivity and/or spatial resolution.
However, a spatially resolved mapping of high-redshift galaxies with the rest-frame $UVJ$ technique should soon become possible with the James Webb Space Telescope (JWST). From the ground, adaptive-optics assisted H$\alpha$ mapping of a large sample of star-forming galaxies would be very valuable in this respect,
being potentially able to separate non-H$\alpha$ emitting bulges from actively star-forming disks. Ultimately, if a tension would persist between predicted and observed quenched fractions, this would indicate that the SFRD/sSFR ratio was observationally underestimated, e.g., at high redshifts  the sSFR of MS galaxies as from Equation \eqref{eq:ssfr} was overestimated and/or  the SFR-$M_*$  relation of star-forming disks may deviate from linearity. 
For example, if one were to insist for the crossover of the quenched and star forming fractions to take place at $z=1$ rather than at $z=3$, then 
from Figure \ref{rho} one sees that this can be achieved for $\eta\simeq 0.60$. But given the good match to the data at $z=0 $ with $\eta=1$, it would be the time evolution of the main sequence that would have to be changed in such a way to have  its sSFR reduced by $\sim 40\%$ by $z=1$, over the rate given by Equation \eqref{eq:ssfr}.

One prediction of these simple calculations is that there should be a great deal of {\it quenched stellar mass} beyond $z\sim 2$, actually exceeding the star-forming mass. At first sight this may appear to be at variance with the \cite{Peng10} {\it mass quenching} and {\it environment quenching} paradigm, as at very high redshift galaxies may have not reached yet the Schechter cutoff mass $M^*$, above which mass quenching efficiently operates, while environmental overdensities have not yet grown enough to trigger environmental (satellite)  quenching. However, in the local Universe there are various examples of early quenching. Globular clusters are the oldest objects with a well measured age. They were formed and quenched beyond $z\sim 3$ and their progenitors may have been ten or more times more massive than the surviving clusters (e.g., \citealt{renzini15b}, and references therein). Local ultra-faint dwarf galaxies appear to be as old as globular clusters and like them were soon quenched as well \citep{brown14}, while most may have been (tidally) destroyed. So, {\it infant mortality} appears to have operated quite effectively already at very early times and on mass scales of  $\sim 10^7-10^8\,\msun$ objects, or less, which may account for up to a few percent of the stellar mass at $z=0$, not too far from the mass of the  quenched fraction for $t\lsim 2$ Gyr seen in Figure \ref{rho}.

\section{Conclusions}

In this paper only two observationally derived relations have been used, namely, the evolution of the SFRD since $z\simeq 8$ as compiled and fit by \cite{md14} and the evolution of the sSFR for near main sequence  galaxies as analytically rendered by \cite{Peng10}. Taking these two relations for granted, their combination has several relevant consequences for the evolution and  build up of stellar mass and star formation quenching on a global scale.The main implications of this simple reading of the cosmic star formation history are the following.

\begin{itemize}
\item
Quenching of star formation within galaxies must start at very early times, and by cosmic time $t\sim 2$ Gyr on ($z\lsim 3$) the stellar mass in the {\it quenched fraction}  can dominate over the star-forming fraction. By star-forming fraction one means those galaxies, or part of them,  whose sSFR  follows the MS relation as given by Equation \eqref{eq:ssfr},
i.e., sSFR $\simeq$ sSFR$_{\rm MS}$. By quenched fraction one means all galaxies for which sSFR $\ll$ sSFR$_{\rm MS}$,  including the parts of star-forming galaxies with sSFR $\ll$ sSFR$_{\rm MS}$, such as quenched bulges and stellar halos.

\item
With the adopted runs of the SFRD and the sSFR,  by $z\sim 0$ only $\sim 22\%$ of the stellar mass is contained in star-forming environments, i.e., in star-forming disks, with $\sim 78\%$ being instead in quenched galaxies or in the quenched bulges and stellar halos of galaxies still supporting star formation in their disk. This appears to be in fair agreement with the corresponding observed fractions.

\item
The run of the sSFR of star-forming galaxies is observationally more uncertain for $z\gsim 2$. It is argued that a modest increase beyond $z=2$ (say, by a factor of $\sim 2$) minimizes the need for quenching at very high redshift  while offering a better  match to the observed  stellar mass density as reported  by \cite{md14}. Instead, a steeper increase of the sSFR (such as in \citealt{stark13} or \citealt{salmon15}) would require an extremely high quenching rate already at $z\sim 8$, almost equal to the SFRD.

\item 
Over the last $\sim 8$ Gyr the stellar mass in star-forming environments (star-forming disks) has remained nearly constant, meaning that the observed drop in the cosmic SFRD over the same time interval is due almost entirely to the secular decrease of the sSFR of main sequence galaxies. This means that during such time interval the SFRD was almost precisely compensated by the quenching rate density, i.e., as much new stellar mass is produced as it gets quenched,
quite a remarkable {\it coincidence} indeed. Hence, both quenching and the decline of sSFR$_{\rm MS}$ concur in establishing the observed decline of the SFRD from $z=2$ to 0. 

\item

There may be a moderate tension between the predicted and observed quenched fractions at $z\gsim 2$, but existing observational data are still insufficient to fully test much beyond the local Universe the consequences of Equations \eqref{eq:sfrd} and \eqref{eq:ssfr}  combined, as explored in this paper. This would require a full recovery of quenched galaxies and quenched spheroids all the way to high redshifts, decomposing star-forming {\it disks} from quenched {\it spheroids}, which will  become possible with JWST.

\end{itemize}

Of course, it remains to be understood what are the physical mechanisms of quenching, from the early to the late cosmic times.

\vskip - 0.9 cm
\section*{Acknowledgments}
    
I wish to thank Marcella Carollo,  Emanuele Daddi, Alberto Franceschini, Simon Lilly, Chiara Mancini, Ying-jie Peng  and Giulia Rodighiero for stimulating discussions on these matters. I am indebted to L.E. Abramson for having provided to me his estimate of the quenched and star-forming fractions in the local Universe and for useful comments on an early version of this paper. I am grateful to the NAOJ Subaru Telescope Headquarters for its support and hospitality while the final version of this paper was set up. Funding support from a INAF-PRIN-2014 grant is also acknowledged.


   \end{document}